\documentclass[conference]{IEEEtran}
\IEEEoverridecommandlockouts

\usepackage{cite}
\usepackage{amsmath,amssymb,amsfonts}
\usepackage{algorithmic}
\usepackage{graphicx}
\usepackage{textcomp}
\usepackage{xcolor}

\usepackage{subfigure}
\usepackage{multirow}
\usepackage{makecell}
\usepackage{latexsym}
\usepackage{amsthm}
\usepackage{booktabs}
\usepackage{enumitem}
\usepackage[normalem]{ulem}  %
\usepackage[ruled,vlined,linesnumbered]{algorithm2e}
\usepackage{tabularx}
\usepackage{tikz}
\usepackage{eso-pic}

\def\BibTeX{{\rm B\kern-.05em{\sc i\kern-.025em b}\kern-.08em
    T\kern-.1667em\lower.7ex\hbox{E}\kern-.125emX}}
\begin{document}

\title{
Energy-Aware Wind-Resilient Routing for Truck-Assisted Multi-UAV Delivery under Wind Uncertainty
\\
}

\author{Tianshun Li$^{1}$, 
Yanggang Sheng$^{1}$,  Hongliang Lu$^{2}$, Zhongzhen Wang$^{1}$,
Haoang Li$^{3}$ and Xinhu Zheng$^{3, \ast}$ 
\thanks{$^\ast$ Corresponding author.}
\thanks{$^{1}$Tianshun Li, Yanggang Sheng, and  Zhongzhen Wang are with
        The Hong Kong University of Science and Technology (Guangzhou), China
        {\tt\small tli449@connect.hkust-gz.edu.cn, yanggangs@hkust-gz.edu.cn, wzz1011@hotmail.com}}%
        \thanks{$^{2}$Hongliang Lu is with the Southern University of Science and Technology, China
        {\tt\small honglianglu36@gmail.com}}
        \thanks{$^{3}$Haoang Li and Xinhu Zheng are with the Intelligent Transportation Thrust, Systems Hub, The Hong Kong University of Science and Technology (Guangzhou), China
        {\tt\small haoangli@hkust-gz.edu.cn, xinhuzheng@hkust-gz.edu.cn}}
        }%


\maketitle

\begin{abstract}
Energy feasibility under wind uncertainty is a critical safety issue for low-altitude air-ground delivery. In truck-UAV systems, UAVs complete assigned deliveries and safely return to a mobile truck or depot, while wind-induced propulsion costs vary online and are only partially observable. Existing routing methods often rely on static or deterministic energy models, which may underestimate headwind, crosswind, battery-voltage, and return-feasibility risks. This paper proposes Energy-Aware Wind-Resilient Routing (EWR), an online risk-sensitive planning framework for wind-aware and energy-safe UAV routing. The delivery environment is represented as a time-dependent directed energy graph whose edge costs are updated using delayed noisy wind estimates, payload states, and conservative uncertainty margins. Experiments using synthetic delivery graphs with replayed wind logs from a public truck-UAV delivery dataset show that EWR improves mission success rates and reduces wind-induced return failures.
\end{abstract}

\section{Introduction}

Air-ground delivery has emerged as a promising paradigm for improving the efficiency and flexibility of last-mile logistics~\cite{ICDE}. Ground vehicles provide long-range mobility and payload-carrying capacity, while unmanned aerial vehicles (UAVs) are deployed to serve spatially dispersed customers with high mobility and reduced travel time~\cite{Lei2026HierarchicalRL,gao2026tridelivercooperativeairgroundinstant}. 
However, the practical deployment of air-ground delivery still faces a fundamental safety challenge: UAVs must not only complete assigned delivery tasks but also retain sufficient energy to safely return to the truck or depot under uncertain environmental conditions.

Wind is widely recognized as a critical factor for UAV energy states in low-altitude environments~\cite{WindFieldModeling,EAMC}. 
Learning-based approaches have been developed to predict local wind flow fields from terrain, sparse measurements, or onboard perception, enabling UAVs to anticipate hazardous wind regions before execution~\cite{chenphysics,SAFELEARNING}. Other studies incorporate wind information into safe perception and planning frameworks, allowing quadrotors to select more stable and energy-efficient trajectories in cluttered or turbulent environments~\cite{End2end}. Wind field modeling has also been investigated for multi-drone formation planning, where spatially varying wind affects trajectory feasibility~\cite{shrivastava2026real}. These works demonstrate that wind is not merely an external disturbance but a planning-relevant environmental state.

However, most existing wind-aware UAV navigation studies are designed for single-vehicle trajectory planning, local obstacle avoidance, or wind prediction itself~\cite{duan2024energy,ding2026perception}. They rarely address the coupled decision-making structure of air-ground delivery, where UAV routing, payload state, residual battery, and safe return requirements are tightly interconnected. Moreover, online wind estimates are often delayed and noisy due to sensor limitations, body vibrations, and finite update frequency. Purely reactive replanning can update trajectories when new wind information becomes available, but it usually trusts the current estimate and does not explicitly verify whether the UAV can still safely return to the truck or depot after each routing decision~\cite{RALTIME}. Therefore, there remains a lack of online routing methods that jointly consider wind-aware energy prediction, uncertainty margins, and return-safety constraints in dynamic windy environments.

To address these limitations, this paper proposes an Energy-Aware Wind-Resilient Routing (EWR) framework for wind-aware and energy-safe UAV routing. 
Our contributions can be summarized as:
\begin{enumerate}
    \item This paper constructs a time-varying energy consumption graph model accommodating wind uncertainty. This approach can characterize the impact of tailwinds, headwinds, and crosswinds on the accessibility and energy consumption of UAV segments, providing a unified graph optimization foundation for wind-aware path planning.
    \item An Energy-Aware Wind-Resilient Routing (EWR) online risk-sensitive planning framework is proposed. EWR continuously evaluates segment energy consumption, return-to-home feasibility, and safety margins during mission execution. It not only considers the energy optimality of the current forward path but also explicitly constrains whether a UAV can safely return to the truck or warehouse.
     \item Experiments on synthetic delivery graphs with log-replay wind data show that EWR improves mission success rates and reduces return failures, highlighting the potential benefits of online energy-feasibility checking in simulated wind-uncertain truck-UAV delivery scenarios.
\end{enumerate}

\begin{figure*}
    \centering
    \includegraphics[width=0.8\linewidth]{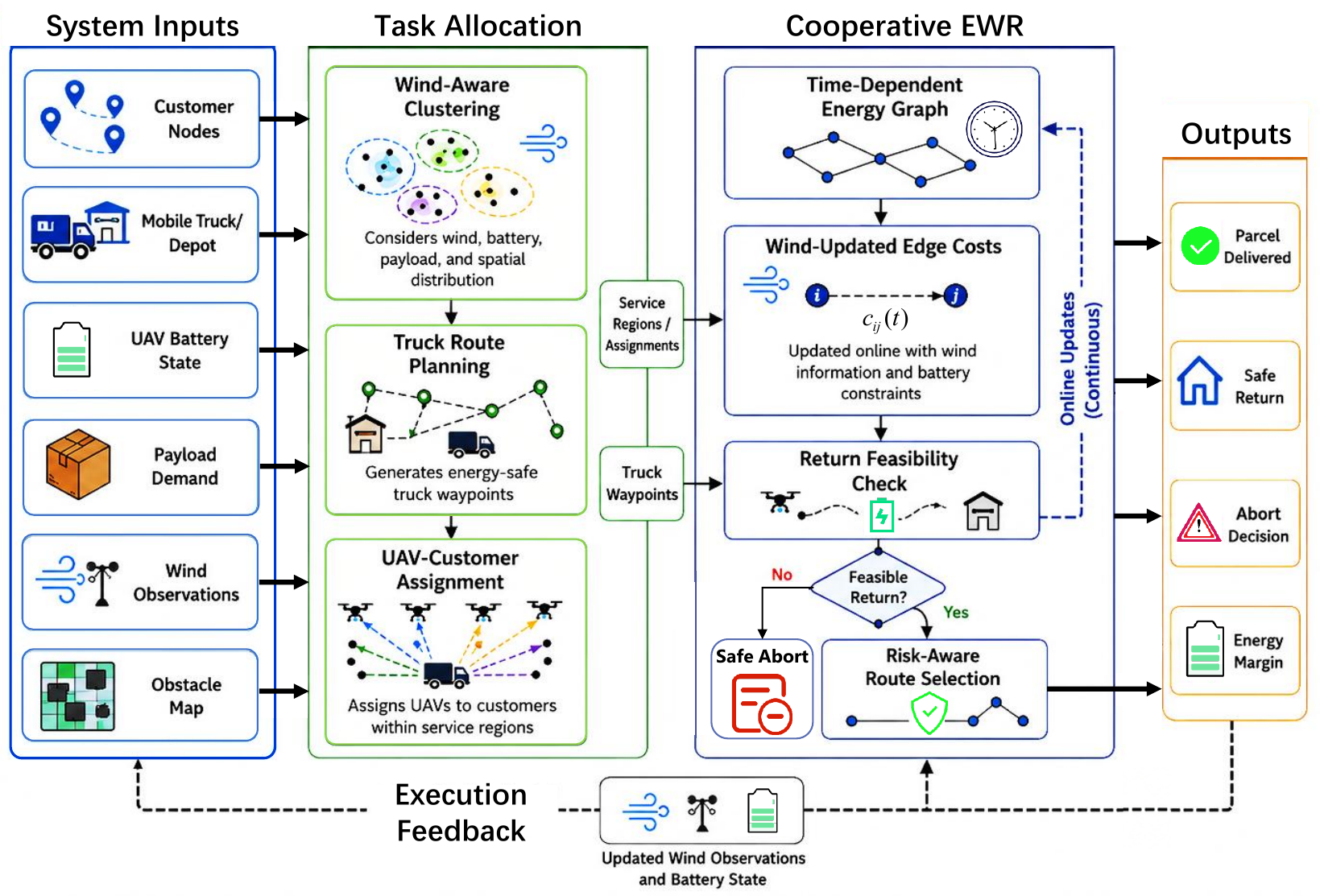}
    \caption{Overview of the proposed framework. }
    \label{fig:framework}
\end{figure*}

\section{Related Work}

\subsection{Energy-Aware UAV Routing}

Recent studies introduced aerodynamic or empirically calibrated energy models for fixed-wing and rotary-wing UAVs, showing that minimum-distance routes are not necessarily minimum-energy routes~\cite{liu2017power,abeywickrama2018comprehensive}. In drone-delivery scenarios, energy constraints have been incorporated into VRP, TSP-D, and truck-drone routing formulations, where payload-dependent consumption, battery capacity, and launch-recovery timing are jointly optimized~\cite{dorling2016vehicle,diller2023energy}. Learning-based and online methods further consider adaptive speed control, power-aware coverage, and battery-aware scheduling under changing mission conditions~\cite{theile2020uav,EAMC,seewald2022energy}.

\subsection{Routing on Dynamic and Time-Dependent Graphs}

Lifelong Planning A* reuses previous search trees when edge costs change~\cite{teng2024reliable}, and D* Lite adapts this idea to robot navigation in unknown terrain~\cite{le2017d}. Anytime Dynamic A* further balances bounded suboptimality and replanning speed, enabling real-time operation when the map changes during execution ~\cite{yeh2025safe}. Surveys of dynamic shortest-path algorithms emphasize the computational benefit of updating shortest paths rather than recomputing from scratch~\cite{martin2017dynamic}. Dynamic and time-dependent graph algorithms offer important tools for efficient replanning, but they rarely model edge weights as uncertain estimates with bounded error and sensing delay. To bridge this gap, we formulate a time-varying energy graph for wind-uncertain air-ground multi-UAV delivery and introduce an online risk-sensitive routing mechanism.

\section{Method}

\subsection{Notation}
Let $\mathcal{C}$ denote the set of customers, with $\mathcal{C}_T$ and $\mathcal{C}_U$ representing truck-served and UAV-served customer sets, respectively. The set of UAVs is denoted by $\mathcal{D}$. We use $\mathcal{G}_t=(\mathcal{V},\mathcal{E},c_t)$ to denote the time-dependent directed energy graph at decision time $t$, where $\mathcal{V}$ is the waypoint set, $\mathcal{E}$ is the directed edge set, and $c_t(e)$ is the online edge cost of $e\in\mathcal{E}$. For a UAV $i\in\mathcal{D}$, $E_i^{\rm rem}(t)$ denotes the remaining battery energy, $E_t(e)$ denotes the predicted traversal energy of edge $e$, and $E_t^{\rm ret}(u)$ denotes the minimum return energy from waypoint $u$ to the truck or depot. The adaptive safety margin is denoted by $E_{\rm safe}(t,e)$ rather than a fixed constant. 
The true local wind is denoted by $\mathbf{W}(t,e)$, whereas $\hat{\mathbf{W}}_t(e)=\hat{\mathbf{W}}(t-\tau_e,e)$ denotes the delayed online wind estimate available to the planner. The bounded estimation error is denoted by $\Delta\mathbf{W}_t(e)$ with confidence radius of $\epsilon_W(t,e)$. Bold symbols, such as $\mathbf{V}^A$ and $\mathbf{W}$, represent vectors, while non-bold speed symbols, such as $V_A$ and $V_G$, represent scalar speeds or path-aligned components.

\subsection{The Relative Wind Model}

For a planned path segment, let $\mathbf{u}_{\eta}$ be the unit tangent direction of the segment. We model the ground velocity as the vector sum of air-relative velocity and local wind.
In practical onboard deployment, the true wind $\mathbf{W}(t,e)$ along an edge is not exactly known at the decision instant. Local wind inference is affected by sensor noise, body vibrations, and finite update frequency. Therefore, EWR uses the delayed online estimate $\hat{\mathbf{W}}_t(e)=\hat{\mathbf{W}}(t-\tau_e,e)$, where $\tau_e\ge 0$ denotes the sensing and update delay on edge $e$. The mismatch between the true wind and the estimate is modeled as
\begin{align}
\mathbf{W}(t,e)=\hat{\mathbf{W}}_t(e)+\Delta\mathbf{W}_t(e),
\qquad
\|\Delta\mathbf{W}_t(e)\|\le \epsilon_W(t,e),
\label{eq:wind_estimation_error}
\end{align}
where $\epsilon_W(t,e)$ is a confidence radius determined by the wind-estimation quality, sensor noise, and elapsed time since the last reliable update. 

For a candidate segment, we decompose the delayed wind estimate into the components parallel and perpendicular to the planned path direction:
\begin{align}
\hat{W}_{\parallel}(t,e) &= \hat{\mathbf{W}}_t(e)^{\top}\mathbf{u}_{\eta}, \\
\hat{\mathbf{W}}_{\perp}(t,e) &= \hat{\mathbf{W}}_t(e)-\hat{W}_{\parallel}(t,e)\mathbf{u}_{\eta},
\end{align}
where \(\mathbf{u}_{\eta}\) denotes the unit tangent direction of the planned ground path. 
We denote by $\hat V_G(t,e)$ the estimated path-aligned ground speed
obtained from the delayed wind estimate and nominal airspeed model.
It should be noted that this relation is used only as a tractable reachability approximation. 
For a multi-rotor UAV, the perpendicular wind component cannot be regarded as energy-neutral, because lateral wind rejection requires additional attitude adjustment and thrust allocation. To avoid falsely accepting a near-critical edge as safe due to wind-estimation error, EWR applies a conservative reachability filter:
\begin{equation}
\left\{
\begin{aligned}
\|\hat{\mathbf{W}}_{\perp}\|+\epsilon_W(t,e) 
&\le \rho_{\perp} V_A, \\
\hat{V}_G-\epsilon_v^W(t,e) 
&> 0 .
\end{aligned}
\right.
\label{eq:wind_feasibility}
\end{equation}
where \(\rho_{\perp}\in(0,1]\) is a lateral control margin and \(\epsilon_v^W(t,e)\) is the induced worst-case ground-speed error bound. A smaller \(\rho_{\perp}\) or a larger \(\epsilon_W(t,e)\) corresponds to a more conservative feasibility criterion under crosswind. In implementation, $\epsilon_v^W(t,e)$ is obtained by projecting the
wind-estimation confidence radius onto the path direction and is
upper-bounded by $\epsilon_W(t,e)$.

To account for the additional energy required by crosswind compensation, the perpendicular wind component is further incorporated into the edge energy model. Specifically, the predicted energy consumption of an edge $e$ for a flyable segment with length $\ell(e)$ is written as
\begin{equation}
E_t(e)=
\frac{\ell(e)}{\hat{V}_G(t,e)}
\left[
P_0+P_{\parallel}(t,e)+P_{\perp}(t,e)+P_{\mathrm{load}}(e)
\right],
\label{eq:edge_energy}
\end{equation}
where \(P_0\) is the nominal cruise power, \(P_{\parallel}(t,e)\) captures the propulsion variation caused by headwind or tailwind, \(P_{\rm load}(e)\) accounts for the payload-induced power increase, and \(P_{\perp}(t,e)\) denotes the additional power required for lateral wind rejection.

\subsection{Task Allocation and EWR Routing}
Multi-UAV coordination is manifested solely through the shared mobile platform and common safe-return constraints; the route planning for each UAV is conducted in a decoupled manner.
As illustrated in Fig.~\ref{fig:framework}, routing is performed on a time-dependent directed energy graph $\mathcal{G}_t=(\mathcal{V},\mathcal{E},c_t)$. Each vertex $v\in\mathcal{V}$ denotes a feasible waypoint, and each directed edge $e=(v,u)\in\mathcal{E}$ denotes a flyable segment with length $\ell(e)$. The local wind estimate along edge $e$ at decision time $t$ is denoted by $\hat{\mathbf{W}}_t(e)$. The online edge cost is defined as
\begin{align}
c_t(e)=E_t(e)+\lambda U_t(e),
\label{eq:ber_edge_cost}
\end{align}
where $E_t(e)$ is the predicted traversal energy, $U_t(e)$ is a wind-induced uncertainty penalty, and $\lambda\ge 0$ controls the risk sensitivity. The energy term $E_t(e)$ is updated using the delayed wind estimate, payload state, and path length. The uncertainty penalty is instantiated as
\begin{align}
U_t(e)=\beta_W\epsilon_W(t,e)+\beta_\tau \tau_e
+\beta_g\|\hat{\mathbf{W}}_t(e)-\hat{\mathbf{W}}_{t^-}(e)\|,
\label{eq:wind_uncertainty_penalty}
\end{align}
where $\hat{\mathbf{W}}_{t^-}(e)$ is the previous accepted estimate and $\beta_W$, $\beta_\tau$, and $\beta_g$ are non-negative weights. This term penalizes edges whose energy prediction is unreliable under partially observed wind, stale updates, or abrupt local changes.

To avoid relying on a fixed energy reserve, EWR uses an adaptive safety margin for each candidate edge $e$:
\begin{align}
E_{\rm safe}(t,e)
=
E_{\rm res}
+\alpha_w\sigma_E(t,e)
+\alpha_b\Delta E_i^{\rm soc}(t)
+\alpha_p\rho_{\rm phase}(t),
\label{eq:adaptive_margin}
\end{align}
where $E_{\rm res}$ is the minimum emergency reserve, $\sigma_E(t,e)$ is the estimated standard deviation of the wind-aware energy prediction, and $\Delta E_i^{\rm soc}(t)$ is the bounded energy-equivalent error induced by state-of-charge estimation uncertainty. The scalar $\rho_{\rm phase}(t)$ is a normalized mission-phase risk indicator, for example distinguishing loaded outbound flight, unloaded return, and final approach. The coefficients $\alpha_w$, $\alpha_b$, and $\alpha_p$ control the sensitivities to wind uncertainty, battery state uncertainty, and mission phase, respectively. The wind-induced energy uncertainty \(\Delta E_W(t,e)\) can be conservatively upper-bounded by
\begin{align}
\Delta E_W(t,e)=L_E(e)\epsilon_W(t,e),
\label{eq:wind_energy_error_bound}
\end{align}
where $L_E(e)$ is a local Lipschitz bound of the edge-energy model with respect to wind perturbations.

\section{Baseline Methods}
First, \textbf{SP-NoWind}~\cite{cheriet2024comparative} denotes a fixed-cost shortest-path baseline based on Dijkstra's shortest-path algorithm, where each edge is weighted only by its geometric length or nominal energy cost without considering wind disturbance. 
Second, \textbf{Energy-SP}~\cite{dorling2016vehicle} is an energy-aware shortest-path method that assigns each edge a distance-dependent energy cost, following the general idea of energy-constrained drone delivery routing, but it does not model time-varying wind. 
Third, \textbf{Initial-Wind-SP} computes wind-aware edge costs using only the wind observation available at mission initialization and then keeps the graph weights fixed during execution. 
Fourth, \textbf{Online-Replan}~\cite{11291475} repeatedly updates edge costs using the latest wind estimate and replans the remaining path with Dijkstra's algorithm whenever new wind information is received. Unlike EWR, it does not incorporate uncertainty margins. 
Fifth, \textbf{D* Lite}~\cite{D*lite} is adopted as a classical incremental dynamic-graph replanning method, which efficiently repairs the shortest path when edge weights change online based on wind-induced edge costs. 
Finally, \textbf{Greedy-Energy} selects at each decision step the locally feasible outgoing edge with the minimum predicted traversal energy.

\section{Experiments}

\subsection{Setup}

Our experiments were implemented in Python on a computer equipped with an AMD Ryzen 7 5800H CPU and an NVIDIA RTX 3050 GPU.
We use a public ATS-based truck-UAV delivery dataset as a source of time-indexed simulated wind logs while constructing controlled synthetic delivery graphs for routing evaluation~\cite{rigoni2022delivery}.
The dataset contains time-indexed wind observations collected during simulated delivery operations, including variations in wind speed and direction over time. These logs are streamed during simulation to update edge costs online, preserving temporally varying wind profiles derived from the dataset logs while maintaining a controllable graph topology.
The normalized initial battery budget is denoted by $B$, where $B=100$ corresponds to full battery capacity and $B=50$ corresponds to a low-budget setting. 
We conduct 100 rounds of experiments for each setting. Within each round, we vary the number of customers and UAVs in the simulation environment. Each UAV is assumed to share identical flight dynamics, cruising speed, and battery endurance. Table.\ref{tab:uav_parameters} represents the parameters of a small rotary-wing delivery UAV and is consistent with empirical UAV energy studies~\cite{EMPRICALVTC}.

\begin{table}[t]
\centering
\caption{Key UAV parameters used in the simulation.}
\label{tab:uav_parameters}
\begin{tabular}{l c c}
\hline
\textbf{Parameter} & \textbf{Symbol} & \textbf{Value} \\
\hline
Nominal airspeed & $V_A$ & $12~\mathrm{m/s}$ \\
Nominal cruise power & $P_0$ & $180~\mathrm{W}$ \\
Battery capacity & $E_{\max}$ & $100~\mathrm{Wh}$ \\
Emergency reserve & $E_{\mathrm{res}}$ & $10~\mathrm{Wh}$ \\
Crosswind margin & $\rho_{\perp}$ & $0.75$ \\
Maximum payload mass & $m_{\mathrm{load}}$ & $0.5~\mathrm{kg}$ \\
\hline
\end{tabular}
\end{table}

\subsection{Evaluation Metrics}

The \textbf{mission success rate} measures the percentage of trials in which all assigned UAV deliveries are completed and the UAV safely returns to the truck or depot:
\begin{equation}
R_{\mathrm{succ}} =
\frac{N_{\mathrm{succ}}}{N_{\mathrm{total}}}\times 100\% ,
\end{equation}
where $N_{\mathrm{succ}}$ and $N_{\mathrm{total}}$ denote the number of successful missions and total test missions, respectively. The \textbf{return failure rate} is defined as
\begin{equation}
R_{\mathrm{fail}} =
\frac{N_{\mathrm{return\_fail}}}{N_{\mathrm{total}}}\times 100\% ,
\end{equation}
where $N_{\mathrm{return\_fail}}$ denotes the number of missions in which a UAV cannot safely return due to insufficient remaining energy. The success rate and return failure rate are not complementary metrics. 
The remaining trials correspond to safe aborts, where the UAVs terminate the mission before entering an unrecoverable return-failure state.
To quantify energy safety during execution, we further report the \textbf{minimum energy margin}:
\begin{equation}
\begin{aligned}
M_{\min} =
\min_t \big[
& E_i^{\mathrm{rem}}(t)
- E_t(e_t)
- E_t^{\mathrm{ret}}(u_t) \\
& - E_{\mathrm{safe}}(t,e_t)
- \Delta E_W(t,e_t)
\big].
\end{aligned}
\end{equation}
A positive $M_{\min}$ indicates that the UAV maintains return feasibility throughout the mission, whereas a negative value implies an energy-unsafe state.

\subsection{Results}
\begin{figure}
    \centering
    \includegraphics[width=1\linewidth]{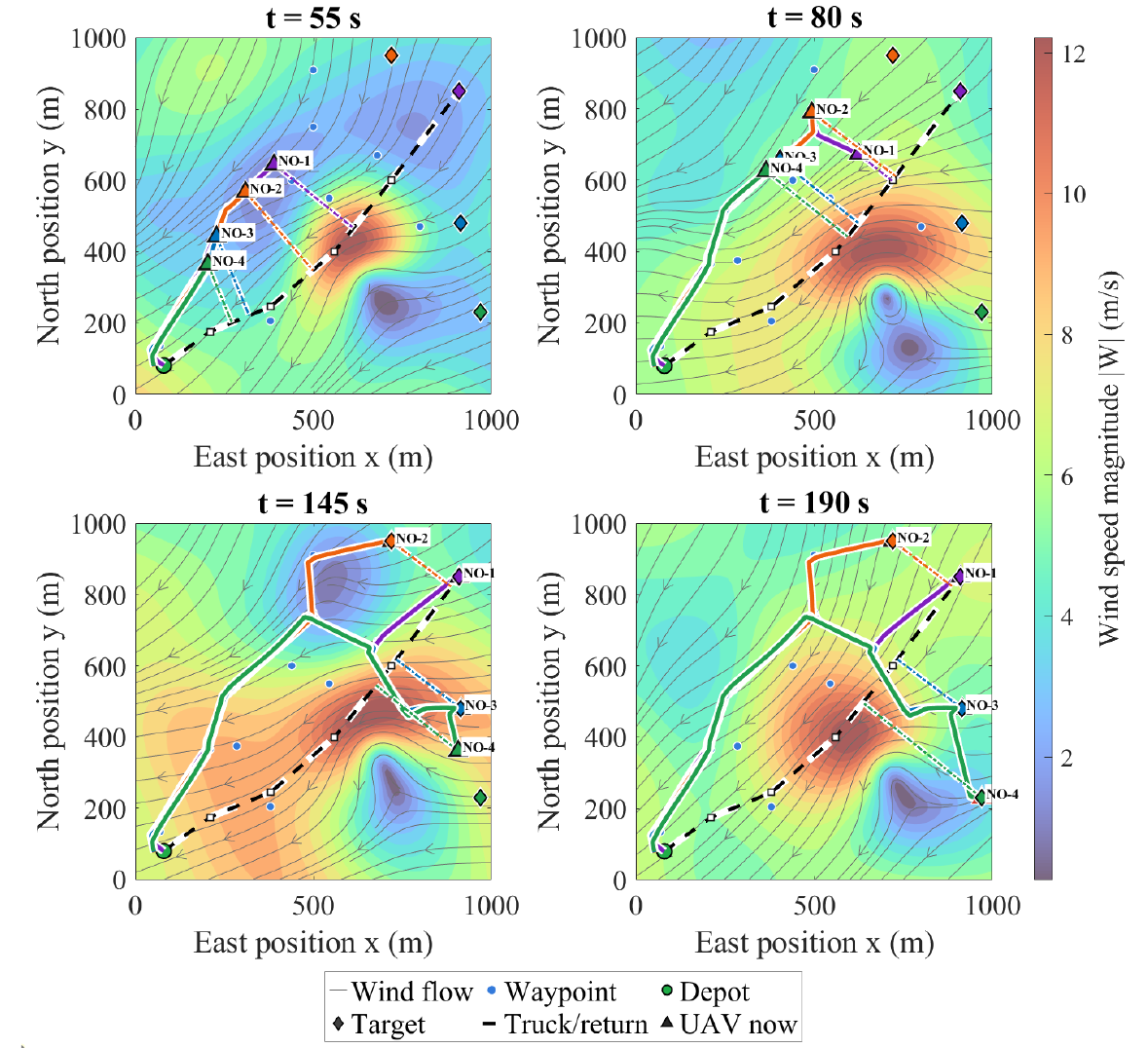}
    \caption{Time-varying routing visualization for the proposed wind-aware energy-safe multi-UAV system. In this experiment, the truck trajectory and UAV-customer assignments are generated before online routing, while EWR updates each UAV route independently using shared truck waypoints and online wind estimates.}
    \label{fig:visual}
\end{figure}

The proposed routing framework that adapts multi-UAV execution under a dynamically evolving wind field is shown in Fig.~\ref{fig:visual}. Across the four snapshots, the UAVs depart from the depot and progressively split toward different task regions, forming a coordinated air-ground delivery pattern. UAV paths tend to bend around high-wind regions and follow comparatively safer corridors, especially near the central red-orange gust zone and the lower-right vortex-like wind structure. The mobile truck route provides a moving ground-side safety reference, and the online return links indicate that each UAV maintains a feasible recovery option during mission execution. 

\begin{figure}
    \centering
    \includegraphics[width=1\linewidth]{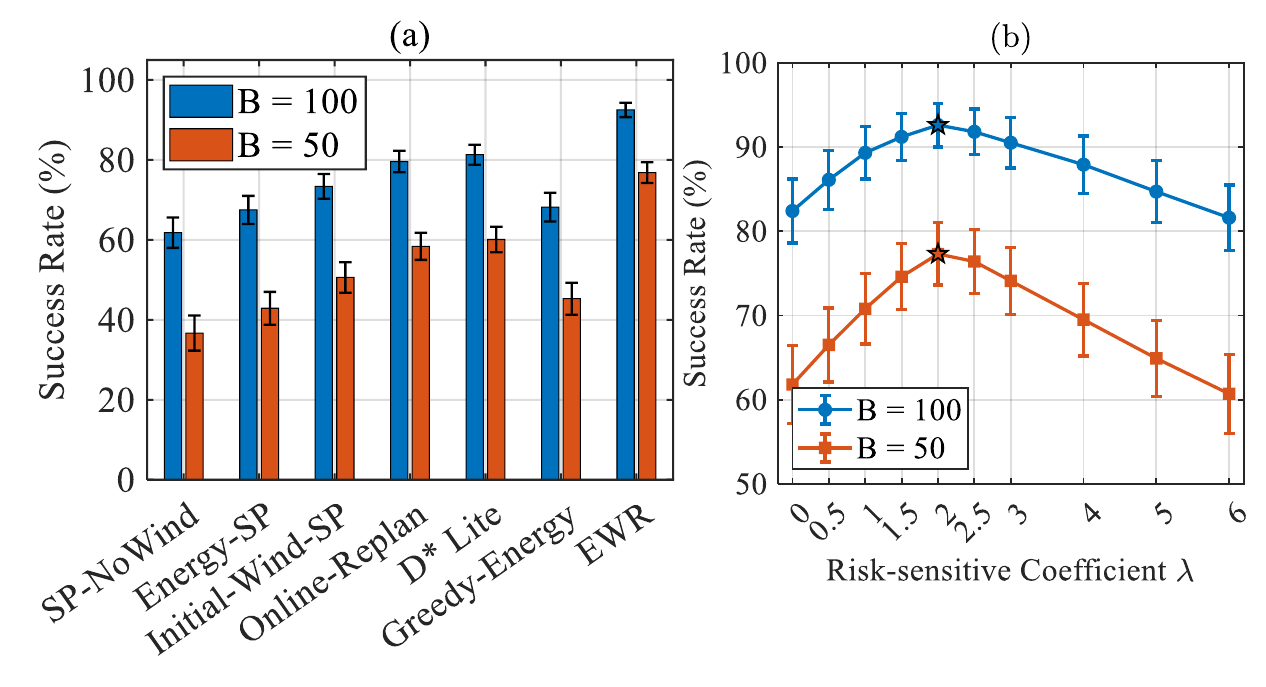}
    \caption{(a) Mission success rate under two battery budgets. (b) Sensitivity analysis of the risk-sensitive coefficient $\lambda$ for EWR under different battery budgets.}
    \label{fig:mission_success}
\end{figure}

Fig.~\ref{fig:mission_success} (a) compares the mission success rates of different routing strategies under two battery budgets. Overall, the proposed EWR method achieves the highest success rate in both battery settings, reaching approximately $92\%$ when $B=100$ and remaining above $75\%$ when the battery budget is reduced to $B=50$. This indicates that EWR can maintain strong mission feasibility even under tighter energy constraints.

When the battery budget decreases from $B=100$ to $B=50$, all methods exhibit a noticeable performance degradation, confirming that limited onboard energy significantly increases the difficulty of completing delivery missions under wind uncertainty. However, the performance drop of EWR is relatively moderate compared with most baselines. In contrast, SP-NoWind and Energy-SP suffer from low success rates under $B=50$, suggesting that path planning based on nominal distance or static energy estimation is insufficient in wind-disturbed environments. Initial-Wind-SP improves over these static baselines by incorporating wind information at the initial planning stage, but its performance remains limited because it cannot adapt to time-varying wind conditions.

\begin{figure}
    \centering
    \includegraphics[width=1\linewidth]{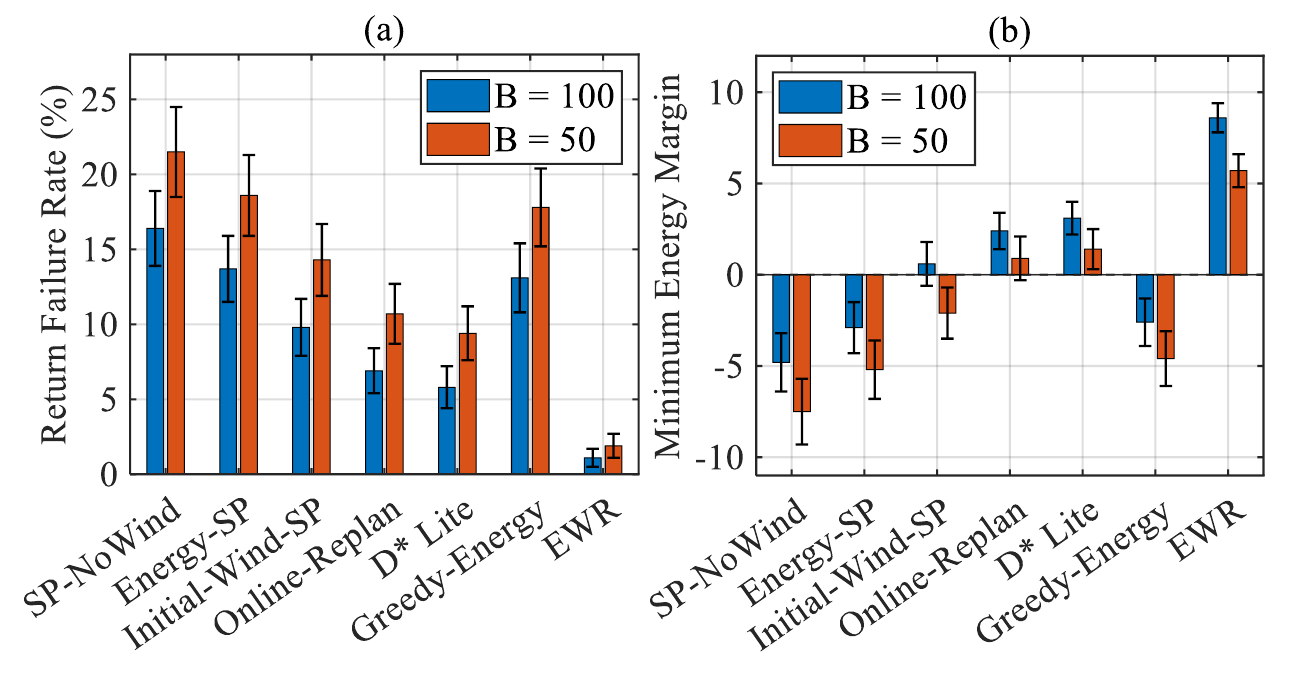}
    \caption{ 
(a) Return safety comparison under wind uncertainty. (b) Minimum residual energy safety margin of different routing baselines under two battery budgets, $B=100$ and $B=50$. }
    \label{fig:return_safety}
\end{figure}

As shown in Fig.~\ref{fig:return_safety}, EWR achieves the most reliable return-safety performance under wind uncertainty. In Fig.~\ref{fig:return_safety} (a), the proposed method consistently yields the lowest return failure rate under both battery budgets. When the battery capacity decreases from $B=100$ to $B=50$, all methods exhibit increased return failures, indicating that tighter energy constraints amplify the adverse impact of wind uncertainty. However, the increase for EWR remains marginal, while conventional shortest-path, greedy, and replanning-based methods suffer from substantially higher failure rates. This demonstrates that merely optimizing nominal path cost or performing reactive replanning is insufficient to guarantee return feasibility in dynamic wind fields.

Fig.~\ref{fig:return_safety} (b) further confirms this observation from the perspective of residual energy safety. Most baseline methods exhibit negative or near-zero minimum safety margins, especially under the lower battery budget, implying that the UAV frequently operates close to or beyond the energy threshold required for safe return. In contrast, EWR maintains a clearly positive safety margin under both $B=100$ and $B=50$, indicating that it preserves sufficient residual energy throughout the mission. The relatively small error bars of EWR also suggest stable performance across repeated trials. 
Overall, results suggest that EWR can reduce return failures in the considered log-replay simulation settings while maintaining a robust energy safety buffer, thereby improving the operational reliability of UAV delivery missions in uncertain wind environments.

\begin{table}[t]
\centering
\caption{Scalability analysis with respect to graph size and the number of UAVs. The values after $\pm$ denote the standard deviation across 100 trials.}
\label{tab:scalability}
\resizebox{\columnwidth}{!}{
\begin{tabular}{c c c c c c}
\hline
\textbf{Customers} & \textbf{UAVs} & \textbf{Nodes} & \textbf{Edges} 
& \textbf{Planning Time} & \textbf{\(R_{\mathrm{succ}}\)} \\
   &  &  &  & \textbf{per Step (ms)} & \textbf{(\%)} \\
\hline
10 & 2 & 42  & 186  & $3.8 \pm 0.6$   & $93.1 \pm 2.4$ \\
20 & 4 & 78  & 412   & $7.6 \pm 1.1$   & $91.7 \pm 2.7$ \\
30 & 6 & 116 & 728   & $14.9 \pm 2.3$  & $89.4 \pm 3.1$ \\
50 & 8 & 184 & 1326 & $31.5 \pm 4.8$  & $86.8 \pm 3.6$ \\
75 & 10 & 268 & 2244  & $64.2 \pm 8.7$ & $83.5 \pm 4.2$ \\
100 & 12 & 352 & 3486  & $158.6 \pm 15.4$ & $80.9 \pm 4.8$ \\
\hline
\end{tabular}
}
\end{table}

Table~\ref{tab:scalability} reports the scalability of EWR under increasing graph sizes and numbers of UAVs. 
As the number of customers increases from 10 to 100, the graph size grows from 42 nodes and 186 directed edges to 352 nodes and 3486 directed edges. 
The average planning time per decision step increases accordingly but remains below 160 ms in the largest tested setting, indicating that the proposed online routing procedure can support real-time replanning in medium-scale delivery scenarios. 
The mission success rate gradually decreases with larger task scales because more UAVs, longer routes, and denser wind-exposed edges increase the probability of energy-critical decisions. 
Nevertheless, EWR maintains a success rate above 80\% even in the largest setting, suggesting reasonable scalability under the tested log-replay wind conditions.

\subsection{Validation with Real-World Wind Logs}

\begin{table}[t]
\centering
\caption{Validation under real-world wind-log replay, $B=100$.}
\label{tab:real_wind_validation}
\begin{tabular}{lccc}
\hline
Method & \(R_{\mathrm{succ}}\) (\%) & \(R_{\mathrm{fail}}\) (\%) & $M_{\min}$ (Wh) \\
\hline
SP-NoWind     & $58.7 \pm 3.9$ & $19.6 \pm 2.8$ & $-6.42 \pm 1.31$ \\
Energy-SP     & $64.3 \pm 3.6$ & $16.8 \pm 2.5$ & $-4.75 \pm 1.18$ \\
Online-Replan & $76.9 \pm 3.1$ & $8.7 \pm 1.9$  & $1.62 \pm 0.74$ \\
D* Lite        & $78.4 \pm 2.9$ & $7.9 \pm 1.7$  & $2.08 \pm 0.69$ \\
EWR            & $\mathbf{86.5 \pm 2.4}$ & $\mathbf{2.8 \pm 0.9}$ & $\mathbf{6.91 \pm 0.86}$ \\
\hline
\end{tabular}
\end{table}

Instead of relying only on the simulated wind traces used in the previous experiments, we replay wind observations from an Automated Surface Observing System (ASOS) weather station~\cite{noaa_asos_1min}. We obtain the selected ASOS wind log through the Iowa
Environmental Mesonet ASOS data interface~\cite{iem_asos_download}. The log contains time-stamped wind speed, wind direction, and gust measurements over a continuous time window.

Table~\ref{tab:real_wind_validation} reports the performance under the real wind-log replay setting. Compared with the nominal shortest-path and online replanning baselines, EWR achieves a higher mission success rate and a lower return failure rate. The improvement is especially evident when the recorded wind log contains abrupt changes in wind direction or gust intensity, where methods without uncertainty margins tend to underestimate the energy required for safe return.

\section{Conclusion}
This work investigates energy-aware routing for truck-UAV delivery missions operating under uncertain wind conditions. 
Experimental results on synthetic delivery graphs and log-replay simulated wind data show that EWR consistently improves mission success rates while maintaining stable behavior across different battery budgets and wind discretizations.
These results indicate that online energy-feasibility checking can serve as a safety layer for low-altitude UAV operations under uncertain wind disturbances. Several limitations remain in this work. First, the model does not account for the prediction of unforeseen sudden strong wind gusts. 
Second, 
the degradation effects of battery aging are not incorporated into the current energy model.

\bibliographystyle{IEEEtran}

\bibliography{references}

@INPROCEEDINGS{WindFieldModeling,
  author={Park, Minhyuk and Au, Tsz-Chiu},
  booktitle={2024 IEEE International Conference on Robotics and Automation (ICRA)}, 
  title={Wind Field Modeling for Formation Planning in Multi-Drone Systems}, 
  year={2024},
  volume={},
  number={},
  pages={12375-12381},
  doi={10.1109/ICRA57147.2024.10610976}}

@INPROCEEDINGS{SAFELEARNING,
  author={Achermann, Florian and Lawrance, Nicholas R. J. and Ranftl, René and Dosovitskiy, Alexey and Chung, Jen Jen and Siegwart, Roland},
  booktitle={2019 International Conference on Robotics and Automation (ICRA)}, 
  title={Learning to Predict the Wind for Safe Aerial Vehicle Planning}, 
  year={2019},
  volume={},
  number={},
  pages={2311-2317},
  doi={10.1109/ICRA.2019.8793547}}

@INPROCEEDINGS{End2end,
  author={Gall, Christian and Fichter, Walter and Ahmad, Aamir},
  booktitle={2024 IEEE International Conference on Robotics and Automation (ICRA)}, 
  title={End-to-End Thermal Updraft Detection and Estimation for Autonomous Soaring Using Temporal Convolutional Networks}, 
  year={2024},
  volume={},
  number={},
  pages={17875-17881},
  doi={10.1109/ICRA57147.2024.10611479}}

@ARTICLE{EAMC,
  author={Datsko, Denys and Nekovar, Frantisek and Penicka, Robert and Saska, Martin},
  journal={IEEE Robotics and Automation Letters}, 
  title={Energy-Aware Multi-UAV Coverage Mission Planning With Optimal Speed of Flight}, 
  year={2024},
  volume={9},
  number={3},
  pages={2893-2900},
  doi={10.1109/LRA.2024.3358581}}

@inproceedings{liu2017power,
  title={A power consumption model for multi-rotor small unmanned aircraft systems},
  author={Liu, Zhilong and Sengupta, Raja and Kurzhanskiy, Alex},
  booktitle={2017 international conference on unmanned aircraft systems (ICUAS)},
  pages={310--315},
  year={2017},
  organization={IEEE}
}

@article{abeywickrama2018comprehensive,
  title={Comprehensive energy consumption model for unmanned aerial vehicles, based on empirical studies of battery performance},
  author={Abeywickrama, Hasini Viranga and Jayawickrama, Beeshanga Abewardana and He, Ying and Dutkiewicz, Eryk},
  journal={IEEE access},
  volume={6},
  pages={58383--58394},
  year={2018},
  publisher={IEEE}
}

@article{dorling2016vehicle,
  title={Vehicle routing problems for drone delivery},
  author={Dorling, Kevin and Heinrichs, Jordan and Messier, Geoffrey G and Magierowski, Sebastian},
  journal={IEEE Transactions on Systems, Man, and Cybernetics: Systems},
  volume={47},
  number={1},
  pages={70--85},
  year={2016},
  publisher={IEEE}
}

@inproceedings{diller2023energy,
  title={Energy-aware uav path planning with adaptive speed},
  author={Diller, Jonathan and Han, Qi},
  booktitle={Proceedings of the 2023 International Conference on Autonomous Agents and Multiagent Systems},
  pages={923--931},
  year={2023}
}

@inproceedings{theile2020uav,
  title={UAV coverage path planning under varying power constraints using deep reinforcement learning},
  author={Theile, Mirco and Bayerlein, Harald and Nai, Richard and Gesbert, David and Caccamo, Marco},
  booktitle={2020 IEEE/RSJ International Conference on Intelligent Robots and Systems (IROS)},
  pages={1444--1449},
  year={2020},
  organization={IEEE}
}

@inproceedings{
chenphysics,
title={Physics-Informed Probabilistic Learning of Low-Altitude Urban Wind from Onboard Motion Data},
author={Minghao Chen},
booktitle={UrbanAI: Harnessing Artificial Intelligence for Smart Cities},
year={2025},
url={https://openreview.net/forum?id=wfYXDYtoYV}
}

@inproceedings{shrivastava2026real,
  title={Real-Time Path Planning for UAVs in Windy Environments Without Computational Fluid Dynamics},
  author={Shrivastava, Abhudaya and Gupta, Shelly and Obradovic, Zoran},
  booktitle={Proceedings of the AAAI Conference on Artificial Intelligence},
  volume={40},
  number={22},
  pages={18540--18548},
  year={2026}
}

@inproceedings{duan2024energy,
  title={Energy-optimized planning in non-uniform wind fields with fixed-wing aerial vehicles},
  author={Duan, Yufei and Achermann, Florian and Lim, Jaeyoung and Siegwart, Roland},
  booktitle={2024 IEEE/RSJ International Conference on Intelligent Robots and Systems (IROS)},
  pages={3116--3122},
  year={2024},
  organization={IEEE}
}

@inproceedings{seewald2022energy,
  title={Energy-aware planning-scheduling for autonomous aerial robots},
  author={Seewald, Adam and de Marina, H{\'e}ctor Garc{\'\i}a and Midtiby, Henrik Skov and Schultz, Ulrik Pagh},
  booktitle={2022 IEEE/RSJ International Conference on Intelligent Robots and Systems (IROS)},
  pages={2946--2953},
  year={2022},
  organization={IEEE}
}

@article{teng2024reliable,
  title={Reliable lifelong planning A*: Technique for re-optimizing reliable shortest paths when travel time distribution updating},
  author={Teng, Wenxin and Chen, Bi Yu},
  journal={Transportation Research Part E: Logistics and Transportation Review},
  volume={188},
  pages={103635},
  year={2024},
  publisher={Elsevier}
}

@inproceedings{D*lite,
author = {Koenig, Sven and Likhachev, Maxim},
title = {D*lite},
year = {2002},
isbn = {0262511290},
publisher = {American Association for Artificial Intelligence},
address = {USA},
booktitle = {Eighteenth National Conference on Artificial Intelligence},
pages = {476–483},
numpages = {8},
location = {Edmonton, Alberta, Canada}
}

@misc{noaa_asos_1min,
  title        = {{1-Minute Page 1 Surface Weather Observations from the Automated Surface Observing System Network}},
  author       = {{NOAA National Centers for Environmental Information}},
  howpublished = {\url{https://www.ncei.noaa.gov/access/metadata/landing-page/bin/iso?id=gov.noaa.ncdc:C00386}},
  note         = {Accessed: 2026-06-19}
}

@misc{iem_asos_download,
  title        = {{Iowa Environmental Mesonet: Download ASOS/AWOS/METAR Data}},
  author       = {{Iowa Environmental Mesonet}},
  howpublished = {\url{https://mesonet.agron.iastate.edu/request/download.phtml}},
  note         = {Accessed: 2026-06-19}
}

@article{ding2026perception,
  title={Perception-Aware Cooperative Path Planning for Multi-UAV Systems in Urban Wind Fields via Deep Reinforcement Learning},
  author={Ding, Jie and Wang, Linshen and Jin, Shuxin and Wang, Di},
  journal={Sensors},
  volume={26},
  number={10},
  pages={2960},
  year={2026},
  publisher={MDPI}
}

@inproceedings{le2017d,
  title={D* lite with reset: Improved version of D* lite for complex environment},
  author={Le, An T and Bui, Minh Q and Le, Than D and Peter, Nauth},
  booktitle={2017 First IEEE International Conference on Robotic Computing (IRC)},
  pages={160--163},
  year={2017},
  organization={IEEE}
}

@inproceedings{yeh2025safe,
  title={Safe Path Planning Based on Anytime Dynamic A-Star Algorithm for Mobile Robots in Indoor Environment},
  author={Yeh, Wei-Ling and Haq, Muhamad Amirul and Ruan, Shanq-Jang},
  booktitle={2025 International Seminar on Intelligent Technology and Its Applications (ISITIA)},
  pages={59--63},
  year={2025},
  organization={IEEE}
}

@article{martin2017dynamic,
  title={Dynamic shortest path and transitive closure algorithms: A survey},
  author={Martin, Daniel P},
  journal={arXiv preprint arXiv:1709.00553},
  year={2017}
}

@misc{gao2026tridelivercooperativeairgroundinstant,
      title={TriDeliver: Cooperative Air-Ground Instant Delivery with UAVs, Couriers, and Crowdsourced Ground Vehicles}, 
      author={Junhui Gao and Yan Pan and Qianru Wang and Wenzhe Hou and Yiqin Deng and Liangliang Jiang and Yuguang Fang},
      year={2026},
      eprint={2604.09049},
      archivePrefix={arXiv},
      primaryClass={cs.RO},
      url={https://arxiv.org/abs/2604.09049}, 
}

@INPROCEEDINGS{11291475,
  author={Nolan, Conor and Cuffe, Paul},
  booktitle={2025 35th Irish Signals and Systems Conference (ISSC)}, 
  title={Shortest Path Planning for Fleets of Delivery Drones: Evaluating the Benefit of Including a Temporal Dimension in the Routing Network}, 
  year={2025},
  volume={},
  number={},
  pages={1-6},
  doi={10.1109/ISSC67739.2025.11291475}}

@inproceedings{cheriet2024comparative,
  title={Comparative analysis of UAV path planning algorithms for efficient navigation in urban 3D environments},
  author={Cheriet, Hichem and Badra, Khellat Kihel and Samira, Chouraqui},
  booktitle={2024 International Conference of the African Federation of Operational Research Societies (AFROS)},
  pages={1--8},
  year={2024},
  organization={IEEE}
}

@INPROCEEDINGS{EMPRICALVTC,
  author={Abeywickrama, Hasini Viranga and Jayawickrama, Beeshanga Abewardana and He, Ying and Dutkiewicz, Eryk},
  booktitle={2018 IEEE 88th Vehicular Technology Conference (VTC-Fall)}, 
  title={Empirical Power Consumption Model for UAVs}, 
  year={2018},
  volume={},
  number={},
  pages={1-5},
  doi={10.1109/VTCFall.2018.8690666}}

@inproceedings{rigoni2022delivery,
  title={Delivery with UAVs: a simulated dataset via ATS},
  author={Rigoni, Giulio and Pinotti, Cristina M and Bhumika and Das, Debasis and Das, Sajal K},
  booktitle={2022 IEEE 95th Vehicular Technology Conference:(VTC2022-Spring)},
  pages={1--6},
  year={2022},
  organization={IEEE}
}

@ARTICLE{RALTIME,
  author={Moon, Brady and Sachdev, Sagar and Yuan, Junbin and Scherer, Sebastian},
  journal={IEEE Robotics and Automation Letters}, 
  title={Time-Optimal Path Planning in a Constant Wind for Uncrewed Aerial Vehicles Using Dubins Set Classification}, 
  year={2024},
  volume={9},
  number={3},
  pages={2176-2183},
  doi={10.1109/LRA.2023.3333167}}

@INPROCEEDINGS{ICDE,
  author={Gao, Junhui and Wang, Qianru and Zhang, Xin and Shi, Juan and Zhao, Xiang and Han, Qingye and Pan, Yan},
  booktitle={2024 IEEE 40th International Conference on Data Engineering (ICDE)}, 
  title={Cooperative Air-Ground Instant Delivery by UAVs and Crowdsourced Taxis}, 
  year={2024},
  volume={},
  number={},
  pages={4153-4166},
  doi={10.1109/ICDE60146.2024.00120}}

@article{Lei2026HierarchicalRL,
  title={Hierarchical Reinforcement Learning for Cooperative Air-Ground Delivery in Urban System},
  author={Songxin Lei and Chun-mei Ma and Haomin Wen and Yexin Li and Lizhenghe Chen and Qianyu Yang and Fugee Tsung and Lei Chen and Sijie Ruan and Yuxuan Liang},
  journal={ArXiv},
  year={2026},
  volume={abs/2602.12913},
  url={https://api.semanticscholar.org/CorpusID:285607462}
}

\end{document}